\documentclass[aps,prl,twocolumn,showpacs,groupedaddress,nofootinbib]{revtex4}  
\usepackage{graphicx}  
\usepackage{dcolumn}   
\usepackage{bm}        
\usepackage{amssymb}   

\newcommand{\mpt}{\slash\kern-5pt p_T}

\begin{document}

\hspace{5.2in} \mbox{FERMILAB-PUB-06-450-E}

\title{Search for techniparticles in $e +$jets events at D0}
%
\author{                                                                      
V.M.~Abazov,$^{35}$                                                           
B.~Abbott,$^{75}$                                                             
M.~Abolins,$^{65}$                                                            
B.S.~Acharya,$^{28}$                                                          
M.~Adams,$^{51}$                                                              
T.~Adams,$^{49}$                                                              
E.~Aguilo,$^{5}$                                                              
S.H.~Ahn,$^{30}$                                                              
M.~Ahsan,$^{59}$                                                              
G.D.~Alexeev,$^{35}$
% ...
\\                                                                            
\vskip 0.30cm                                                                 
\centerline{(D\O\ Collaboration)}                                             
\vskip 0.30cm                                                                 
}                                                                             
\affiliation{                                                                 
\centerline{$^{1}$Universidad de Buenos Aires, Buenos Aires, Argentina}       
\centerline{$^{2}$LAFEX, Centro Brasileiro de Pesquisas F{\'\i}sicas,         
                  Rio de Janeiro, Brazil}                                     
\centerline{$^{3}$Universidade do Estado do Rio de Janeiro,                   
                  Rio de Janeiro, Brazil}                                     
\centerline{$^{4}$Instituto de F\'{\i}sica Te\'orica, Universidade            
                  Estadual Paulista, S\~ao Paulo, Brazil}                     
\centerline{$^{5}$University of Alberta, Edmonton, Alberta, Canada,           
                  Simon Fraser University, Burnaby, British Columbia, Canada,}
\centerline{York University, Toronto, Ontario, Canada, and                    
                  McGill University, Montreal, Quebec, Canada}                
\centerline{$^{6}$University of Science and Technology of China, Hefei,       
                  People's Republic of China}                                 
\centerline{$^{7}$Universidad de los Andes, Bogot\'{a}, Colombia}             
\centerline{$^{8}$Center for Particle Physics, Charles University,            
                  Prague, Czech Republic}                                     
\centerline{$^{9}$Czech Technical University, Prague, Czech Republic}         
\centerline{$^{10}$Center for Particle Physics, Institute of Physics,         
                   Academy of Sciences of the Czech Republic,                 
                   Prague, Czech Republic}                                    
\centerline{$^{11}$Universidad San Francisco de Quito, Quito, Ecuador}        
\centerline{$^{12}$Laboratoire de Physique Corpusculaire, IN2P3-CNRS,         
                   Universit\'e Blaise Pascal, Clermont-Ferrand, France}      
\centerline{$^{13}$Laboratoire de Physique Subatomique et de Cosmologie,      
                   IN2P3-CNRS, Universite de Grenoble 1, Grenoble, France}    
\centerline{$^{14}$CPPM, IN2P3-CNRS, Universit\'e de la M\'editerran\'ee,     
                   Marseille, France}                                         
\centerline{$^{15}$Laboratoire de l'Acc\'el\'erateur Lin\'eaire,              
                   IN2P3-CNRS et Universit\'e Paris-Sud, Orsay, France}       
\centerline{$^{16}$LPNHE, IN2P3-CNRS, Universit\'es Paris VI and VII,         
                   Paris, France}                                             
\centerline{$^{17}$DAPNIA/Service de Physique des Particules, CEA, Saclay,    
                   France}                                                    
\centerline{$^{18}$IPHC, IN2P3-CNRS, Universit\'e Louis Pasteur, Strasbourg,  
                   France, and Universit\'e de Haute Alsace,                  
                   Mulhouse, France}                                          
\centerline{$^{19}$Institut de Physique Nucl\'eaire de Lyon, IN2P3-CNRS,      
                   Universit\'e Claude Bernard, Villeurbanne, France}         
\centerline{$^{20}$III. Physikalisches Institut A, RWTH Aachen,               
                   Aachen, Germany}                                           
\centerline{$^{21}$Physikalisches Institut, Universit{\"a}t Bonn,             
                   Bonn, Germany}                                             
\centerline{$^{22}$Physikalisches Institut, Universit{\"a}t Freiburg,         
                   Freiburg, Germany}                                         
\centerline{$^{23}$Institut f{\"u}r Physik, Universit{\"a}t Mainz,            
                   Mainz, Germany}                                            
\centerline{$^{24}$Ludwig-Maximilians-Universit{\"a}t M{\"u}nchen,            
                   M{\"u}nchen, Germany}                                      
\centerline{$^{25}$Fachbereich Physik, University of Wuppertal,               
                   Wuppertal, Germany}                                        
\centerline{$^{26}$Panjab University, Chandigarh, India}                      
\centerline{$^{27}$Delhi University, Delhi, India}                            
\centerline{$^{28}$Tata Institute of Fundamental Research, Mumbai, India}     
\centerline{$^{29}$University College Dublin, Dublin, Ireland}                
\centerline{$^{30}$Korea Detector Laboratory, Korea University,               
                   Seoul, Korea}                                              
\centerline{$^{31}$SungKyunKwan University, Suwon, Korea}                     
\centerline{$^{32}$CINVESTAV, Mexico City, Mexico}                            
\centerline{$^{33}$FOM-Institute NIKHEF and University of                     
                   Amsterdam/NIKHEF, Amsterdam, The Netherlands}              
\centerline{$^{34}$Radboud University Nijmegen/NIKHEF, Nijmegen, The          
                  Netherlands}                                                
\centerline{$^{35}$Joint Institute for Nuclear Research, Dubna, Russia}       
\centerline{$^{36}$Institute for Theoretical and Experimental Physics,        
                   Moscow, Russia}                                            
\centerline{$^{37}$Moscow State University, Moscow, Russia}                   
\centerline{$^{38}$Institute for High Energy Physics, Protvino, Russia}       
\centerline{$^{39}$Petersburg Nuclear Physics Institute,                      
                   St. Petersburg, Russia}                                    
\centerline{$^{40}$Lund University, Lund, Sweden, Royal Institute of          
                   Technology and Stockholm University, Stockholm,            
                   Sweden, and}                                               
\centerline{Uppsala University, Uppsala, Sweden}                              
\centerline{$^{41}$Physik Institut der Universit{\"a}t Z{\"u}rich,            
                   Z{\"u}rich, Switzerland}                                   
\centerline{$^{42}$Lancaster University, Lancaster, United Kingdom}           
\centerline{$^{43}$Imperial College, London, United Kingdom}                  
\centerline{$^{44}$University of Manchester, Manchester, United Kingdom}      
\centerline{$^{45}$University of Arizona, Tucson, Arizona 85721, USA}         
\centerline{$^{46}$Lawrence Berkeley National Laboratory and University of    
                   California, Berkeley, California 94720, USA}               
\centerline{$^{47}$California State University, Fresno, California 93740, USA}
\centerline{$^{48}$University of California, Riverside, California 92521, USA}
\centerline{$^{49}$Florida State University, Tallahassee, Florida 32306, USA} 
\centerline{$^{50}$Fermi National Accelerator Laboratory,                     
            Batavia, Illinois 60510, USA}                                     
\centerline{$^{51}$University of Illinois at Chicago,                         
            Chicago, Illinois 60607, USA}                                     
\centerline{$^{52}$Northern Illinois University, DeKalb, Illinois 60115, USA} 
\centerline{$^{53}$Northwestern University, Evanston, Illinois 60208, USA}    
\centerline{$^{54}$Indiana University, Bloomington, Indiana 47405, USA}       
\centerline{$^{55}$University of Notre Dame, Notre Dame, Indiana 46556, USA}  
\centerline{$^{56}$Purdue University Calumet, Hammond, Indiana 46323, USA}    
\centerline{$^{57}$Iowa State University, Ames, Iowa 50011, USA}              
\centerline{$^{58}$University of Kansas, Lawrence, Kansas 66045, USA}         
\centerline{$^{59}$Kansas State University, Manhattan, Kansas 66506, USA}     
\centerline{$^{60}$Louisiana Tech University, Ruston, Louisiana 71272, USA}   
\centerline{$^{61}$University of Maryland, College Park, Maryland 20742, USA} 
\centerline{$^{62}$Boston University, Boston, Massachusetts 02215, USA}       
\centerline{$^{63}$Northeastern University, Boston, Massachusetts 02115, USA} 
\centerline{$^{64}$University of Michigan, Ann Arbor, Michigan 48109, USA}    
\centerline{$^{65}$Michigan State University,                                 
            East Lansing, Michigan 48824, USA}                                
\centerline{$^{66}$University of Mississippi,                                 
            University, Mississippi 38677, USA}                               
\centerline{$^{67}$University of Nebraska, Lincoln, Nebraska 68588, USA}      
\centerline{$^{68}$Princeton University, Princeton, New Jersey 08544, USA}    
\centerline{$^{69}$State University of New York, Buffalo, New York 14260, USA}
\centerline{$^{70}$Columbia University, New York, New York 10027, USA}        
\centerline{$^{71}$University of Rochester, Rochester, New York 14627, USA}   
\centerline{$^{72}$State University of New York,                              
            Stony Brook, New York 11794, USA}                                 
\centerline{$^{73}$Brookhaven National Laboratory, Upton, New York 11973, USA}
\centerline{$^{74}$Langston University, Langston, Oklahoma 73050, USA}        
\centerline{$^{75}$University of Oklahoma, Norman, Oklahoma 73019, USA}       
\centerline{$^{76}$Oklahoma State University, Stillwater, Oklahoma 74078, USA}
\centerline{$^{77}$Brown University, Providence, Rhode Island 02912, USA}     
\centerline{$^{78}$University of Texas, Arlington, Texas 76019, USA}          
\centerline{$^{79}$Southern Methodist University, Dallas, Texas 75275, USA}   
\centerline{$^{80}$Rice University, Houston, Texas 77005, USA}                
\centerline{$^{81}$University of Virginia, Charlottesville,                   
            Virginia 22901, USA}                                              
\centerline{$^{82}$University of Washington, Seattle, Washington 98195, USA}  
}                                                                             
\date{Dec 5, 2006}

\begin{abstract}
We search for the technicolor process $p\overline p\rightarrow \rho_T/\omega_T
\rightarrow W \pi_T$ in events containing one electron and two jets,
in data corresponding
to an integrated luminosity of 390 pb$^{-1}$, recorded by
the D0 experiment at the Fermilab Tevatron. 
Technicolor predicts that technipions, $\pi_T$, decay dominantly into
$b\overline{b}$, $b\overline{c}$, or $\overline{b}c$, 
depending on their charge. 
In these events $b$ and $c$ quarks are identified by their
secondary decay vertices within jets.
Two analysis methods based on topological variables are presented. 
Since no excess above the standard model prediction was found,
the result is presented as an exclusion in the $\pi_T$ vs. $\rho_T$ mass plane for 
a given set of model parameters.

\end{abstract}

\pacs{12.60.Nz, 13.85.Rm}
\maketitle

Technicolor (TC), first formulated by Weinberg and Susskind~\cite{Wei1,Sus1},
provides a dynamical explanation of electroweak symmetry breaking through a
new strong $SU(N_{TC})$ gauge interaction acting on new fermions, called
``technifermions.'' Technicolor is a non-Abelian gauge theory modeled after
Quantum Chromodynamics (QCD). In its low-energy limit, a spontaneous breaking
of the  global chiral symmetry in the technifermion sector
leads to electroweak symmetry breaking. The
Nambu-Goldstone bosons produced in this process are called technipions,
$\pi_T$, in analogy with the pions of QCD. Three of these technipions become
the longitudinal components of the $W$ and $Z$ bosons, making them 
massive.

An additional gauge interaction, called extended
technicolor~\cite{etc1,etc2}, couples standard model fermions and
technifermions to provide a mechanism for generating quark and lepton masses. By
limiting the running of the technicolor coupling constant, walking
technicolor~\cite{wtc} avoids flavor-changing neutral currents. 
To generate masses as large as the top quark mass,
another interaction, topcolor, seems to be necessary, thereby 
giving rise to topcolor-assisted technicolor models~\cite{tc2}.

Extensions of the basic technicolor model tend to require the
number $N_D$ of technifermion doublets to be large. In general, the
technicolor scale $\Lambda_{TC}\approx O(1) \times F_{TC}$, where $F_{TC}$
is the technipion decay constant, depends inversely on the number of
technifermion doublets: $F_{TC} \approx 246\,{\rm GeV}/ \sqrt{ N_{D} }$.
For large $N_{D}$, the lowest lying technihadrons have masses on the order of
few hundred GeV. This scenario is referred to as low-scale
technicolor~\cite{Lane1}. Low-scale
technicolor models predict the existence of scalar technimesons, $\pi_T^\pm$ and
$\pi_T^0$, and vector technimesons, $\rho_T$ and $\omega_T$. General features
of low-scale technicolor have been summarized in the technicolor strawman
model (TCSM)~\cite{Lane2,Lane4}. The analysis presented in this paper is based on Ref.~\cite{Lane4}.

Vector   technimesons are expected to be produced with
substantial rates at the Fermilab Tevatron Collider via the
Drell-Yan-like electroweak process $p\overline p\rightarrow \rho_T+X$
or $\omega_T+X$. In walking technicolor, it is expected that vector
technimesons decay to a gauge boson ($\gamma$, $W$, $Z$) and a technipion or
to fermion-antifermion pairs. The production cross sections and branching
fractions depend on the masses of the vector technimesons, $M(\rho_T)$ and
$M(\omega_T)$, on the technicolor-charges of the technifermions, on the
mass differences between the vector and scalar technimesons, which determine
the spectrum of accessible decay channels, and on two mass parameters, $M_A$
for axial-vector and $M_V$ for vector couplings. The parameter $M_V$ controls
the rate for the decay $\rho_T,\omega_T\rightarrow\gamma+\pi_T$ and is
unknown {\it a priori}. Scaling from the QCD decay
$\rho,\omega\rightarrow\gamma+\pi^0$, the authors of Ref.~\cite{Lane4}
suggest a value of several hundred GeV. We set $M_A=M_V$, and
evaluate the production and decay rates at two different values: 100 and 500
GeV.  For all other parameters, we use the default values quoted in Table III
of  Ref.~\cite{Lane4}. 
Technipion coupling to the standard model
particles is proportional to their masses,
therefore technipions in the mass range
considered here predominantly decay into $b\bar{b}$, $b\bar{c}$, or
$\bar{b}c$, depending on their charge.

In this Letter, we describe a search for the decay of vector technimesons to $W\pi_T$,
followed by the decays $W\to e\nu$ and $\pi_T\to b\bar{b}$, $b\bar{c}$, or
$c\bar{b}$. In the D0 detector, which is described in detail in
Ref.~\cite{d0det}, the signature of this process is an isolated electron and
missing transverse momentum~($\mpt$) from the undetected neutrino from the
decay of the $W$ boson, and two jets of hadrons coming from the fragmentation
of the quarks from the decay of the technipion. Jets are reconstructed  
using the Run II cone algorithm~\cite{cone} with a cone size of 0.5. 
We search for events with this signature in the data collected with a single electron
trigger until July 2004 and corresponding to
an integrated luminosity of 388$\pm$25~pb$^{-1}$~\cite{oldlumi}.

There are a number of standard model processes that can result in the same
final state signature as $W\pi_T$ production. Vector boson production in association
with jets is the dominant background. $Z$ boson production can be suppressed by
vetoing on a second electron and requiring significant $\mpt$. Most of
the jets in $W$+jets events originate from the fragmentation of light quarks
or gluons and therefore requiring the explicit identification of at least one
jet from the fragmentation of a $b$ or $c$ quark suppresses most of this background,
leaving only $W+b\bar{b}$, $W+b$, $W+c\bar{c}$, and $W+c$ events. Top quark
production followed by the decay to $e\nu b$ is another background.
Top-antitop quark pair production typically results in either an additional
lepton or a higher jet multiplicity from the decay of the second top quark,
and this background can be reduced by selecting events with exactly two jets. Single top
quark production is an irreducible background, but it has a smaller cross
section. We simulate all these processes using either {\sc
  pythia}~\cite{pythia} or {\sc alpgen}~\cite{alpg} Monte Carlo (MC) generators, 
followed by the D0 detector simulation based on {\sc geant}~\cite{geant}.
Quark hadronization and fragmentation is
simulated using {\sc pythia}.

The multijet background is due to events with
poorly measured jets, resulting in missing momentum and a jet that
is misidentified as an electron. Background from the mistagged $W$+jets process originates
from events in which a light-quark or gluon jet is incorrectly identified as a
$b$ jet. These instrumental background contributions are estimated from the
same data sample before requiring the identification of a $b$ jet. 

\begin{figure*}
\includegraphics[scale=0.29]{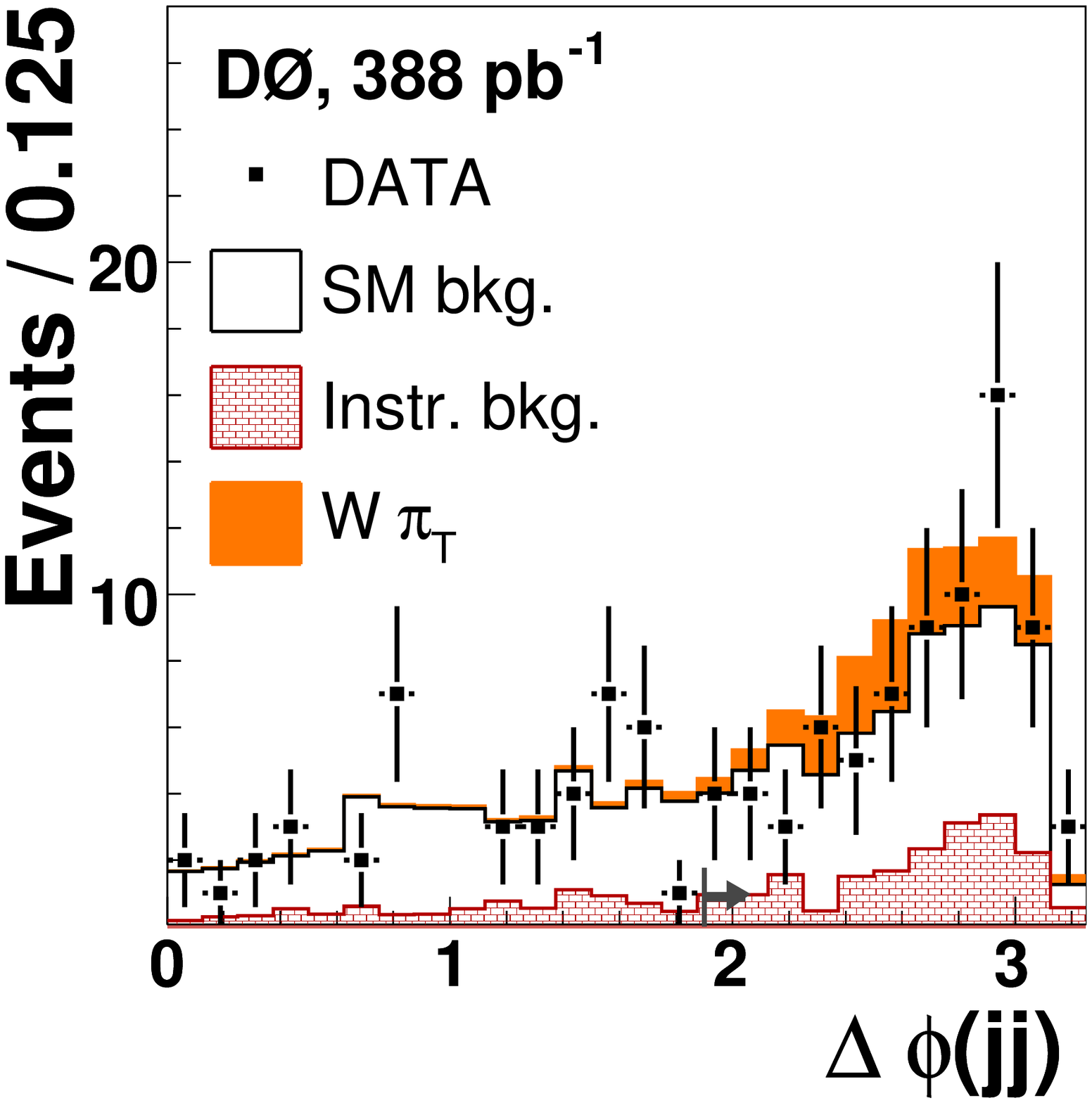}
\includegraphics[scale=0.29]{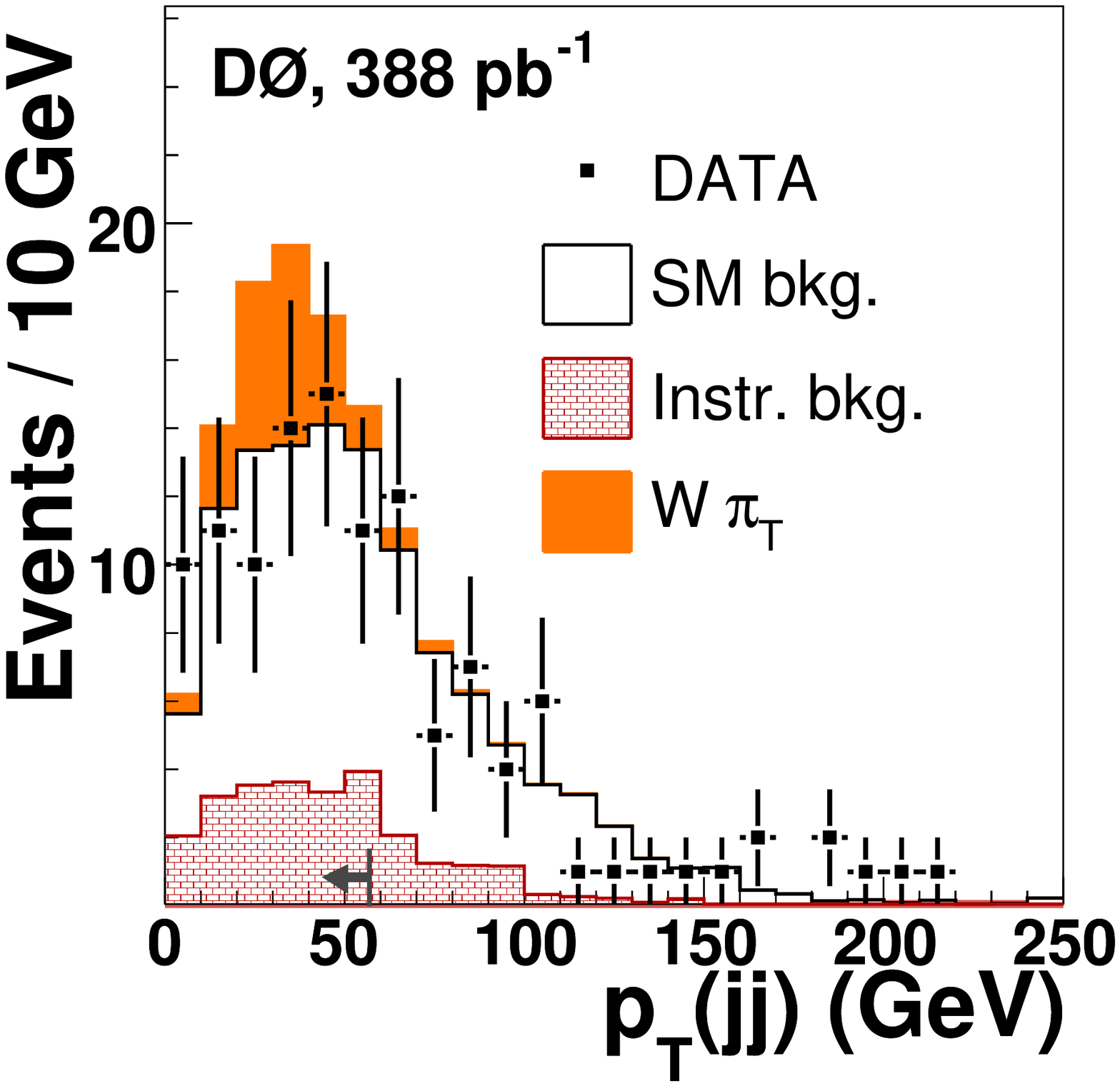}
\includegraphics[scale=0.29]{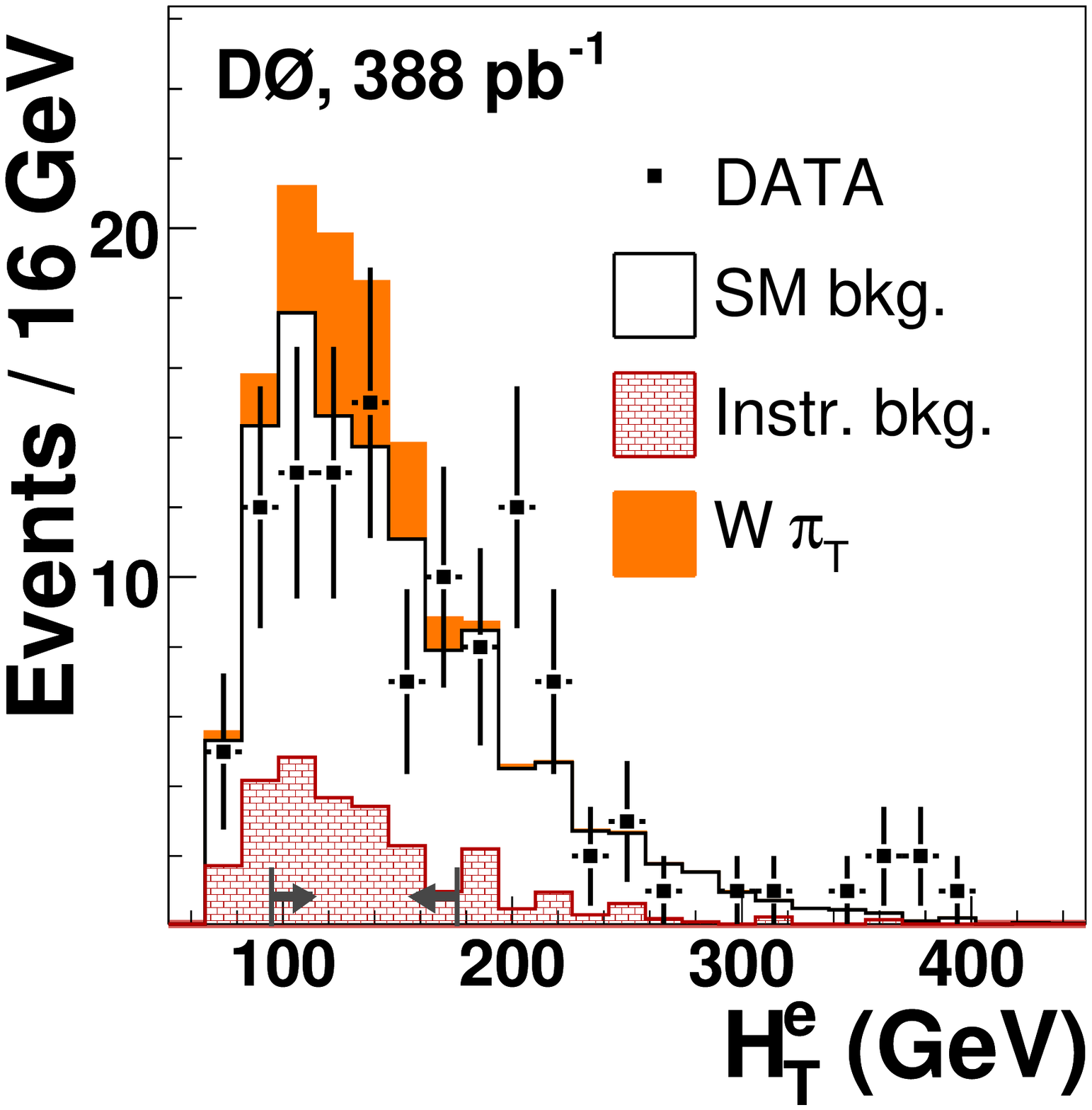}
\caption{\label{fig:wide} Distributions of $\Delta \phi(jj)$, $p_T(jj)$ and $H_T^e$ after final kinematic selection. The $W \pi_{T}$ signal is shown for $M(\rho_T)=210$~GeV and $M(\pi_{T})=110$~GeV. Arrows at 
the bottom indicate the cuts applied in the cut-based analysis for the signal mass point shown. }
\end{figure*}

We select events in which there is exactly one well-identified electron based
on tracking and calorimeter data with transverse momentum $p_T > 20$~GeV and
pseudorapidity $|\eta| < 1.1$~\cite{pseudorapidity}. 
There must be significant $\mpt$, measured
in two ways: $\mpt^{\text{obj}} >20$~GeV computed as the negative sum of the jet
momentum vectors and the electron momentum vector and $\mpt > 20$~GeV which
also includes the calorimeter energy deposit not assigned to the electron or the
jets. We require the transverse mass $M_T(e\nu) >30$~GeV. 
We further require the presence of exactly two jets
with $p_T > 20$~GeV and $|\eta| < 2.5$. 

To further reduce backgrounds, we take advantage of the long lifetime of $b$ flavored
hadrons. Tracks from the decay products of $b$ hadrons may not project back to
the proton-antiproton collision, but have a significant impact
parameter. They can therefore be identified and used to reconstruct the decay
vertex of the $b$ hadron. A jet is tagged as a $b$ jet if there is a secondary decay vertex within $\Delta {\cal R}=\sqrt{(\Delta\eta)^2+(\Delta\phi)^2}<0.5$ of the jet axis. We require at least one of the jets to be $b$-tagged. This leaves us with 117 events in our final data sample.

The expected background event yields are listed in Table~\ref{tab:nevents}.
When estimating these yields, each Monte Carlo event is weighted by the probability
that at least one jet is tagged as a $b$ jet. The tagging probability is
parameterized as a function of jet flavor, jet $p_T$, and $\eta$. The
efficiency of tagging a jet from the fragmentation of a $b$ quark is derived
from collider data which were enriched in their $b$ jet contents by requiring
a muon to be reconstructed within at least one jet to preferentially select
jets with semileptonic $b$ decays.  The probability of tagging a $c$ jet is
derived from the tagging probability for $b$ jets by multiplying by the ratio
of tagging probabilities for $c$ and $b$ jets derived from MC
simulations. We derive the probability to tag a light-quark or gluon jet from
a set of dijet events, corrected for contamination by $c$ and $b$ jets.  The
Monte Carlo events are also weighted by the ratios of jet and electron finding
efficiencies in Monte Carlo and collider data. 
Electron finding efficiencies
are measured in
$Z\rightarrow ee$ events in both data and Monte Carlo.

\begin{table}
\caption{Number of events observed in the data and expected from signal and background sources after the kinematic selection; only statistical errors are reported. For the expected number of signal events quoted we assume $M(\rho_T)=210$~GeV and $M(\pi_T)=110$~GeV. \label{tab:nevents}}

\begin{tabular}{lrcl}
\hline\hline
Final data sample & 117 \\
\hline
\multicolumn{4}{l}{Signal:}\\
$\rho_T/\omega_T\rightarrow W+\pi_T\rightarrow e\nu b\overline b$ ($M_V=100$~GeV) & 11.1 & $\pm$ &  0.1\\
$\rho_T/\omega_T\rightarrow W+\pi_T\rightarrow e\nu b\overline b$ ($M_V=500$~GeV) & 17.1 & $\pm$ &  0.2\\
\hline
\multicolumn{4}{l}{Physics background:}\\
$t\overline t\rightarrow \ell\nu b q\overline q \overline b$ & 7.9 & $\pm$ &  0.5 \\
$t\overline t\rightarrow \ell^+\nu b\ell^-\nu\overline b $ & 14.1 & $\pm$ &  0.3\\
$W^*\rightarrow tb\rightarrow e\nu b \overline b$ or $\tau\nu b \overline b$ & 3.5 & $\pm$ &  0.1 \\
$tqb\rightarrow e\nu b\overline b$ or $\tau\nu b \overline b$ & 4.3 & $\pm$ &  0.1\\
$W(\rightarrow e\nu)+\mbox{heavy flavor}$  & 56.4 & $\pm$ & 4.2\\
$WZ\rightarrow e\nu b\overline b$  & 1.10 & $\pm$ &  0.02 \\
$Z(\rightarrow e^+e^-)$ & 0.5 & $\pm$ &  0.4 \\
$Z(\rightarrow e^+e^-) + b\overline b$ & 0.60 & $\pm$ &  0.03  \\
\hline
\multicolumn{2}{l}{Instrumental background:} \\
multijet events & 16.3 & $\pm$ &  3.2 \\
mistagged $W(\rightarrow e\nu)$ + jets & 10.3 & $\pm$ &  0.3\\
\hline
Total background & 115.1 & $\pm$ &  5.4 \\
\hline\hline
\end{tabular}
\end{table}

We use the {\sc pythia} event generator to simulate signal events,
modeling  initial state and final state radiation, fragmentation, 
and hadronization. To generate $W\pi_T$ signal events for
a range of values of the 
technimeson masses, we use a fast, parameterized detector
simulation that was tuned to reproduce the kinematic distributions and
acceptances from events simulated with the detailed {\sc geant}-based detector
simulation. 
For the cross section calculations, {\sc
  CTEQ5L}~\cite{CTEQ5} parton distribution functions are used. Finally,
as is appropriate for this Drell-Yan-like process, the
cross section is multiplied by a $K$-factor of 1.3 to approximate NLO
contributions to the cross section~\cite{kfac}.  We generate events with $\rho_T$ masses
from 160~GeV to 220~GeV and assume 
$M(\omega_T)=M(\rho_T)$. The $\pi_T$ mass
values start at the kinematic threshold for $W\pi_T$ production at
$M(\pi_T)=M(\rho_T)-M(W)$ and go down to $M(\pi_T)=M(\rho_T)/2 - 5$~GeV where the decay
channel $\rho_T^{\pm(0)} \rightarrow \pi_T^{\pm(0,\pm)} \pi_T^{0(0,\mp)}$ 
is accessible, reducing the branching fraction of $\rho_T^{\pm(0)}
\rightarrow W\pi_T$.

\begin{figure}
\centering
\includegraphics[scale=0.21]{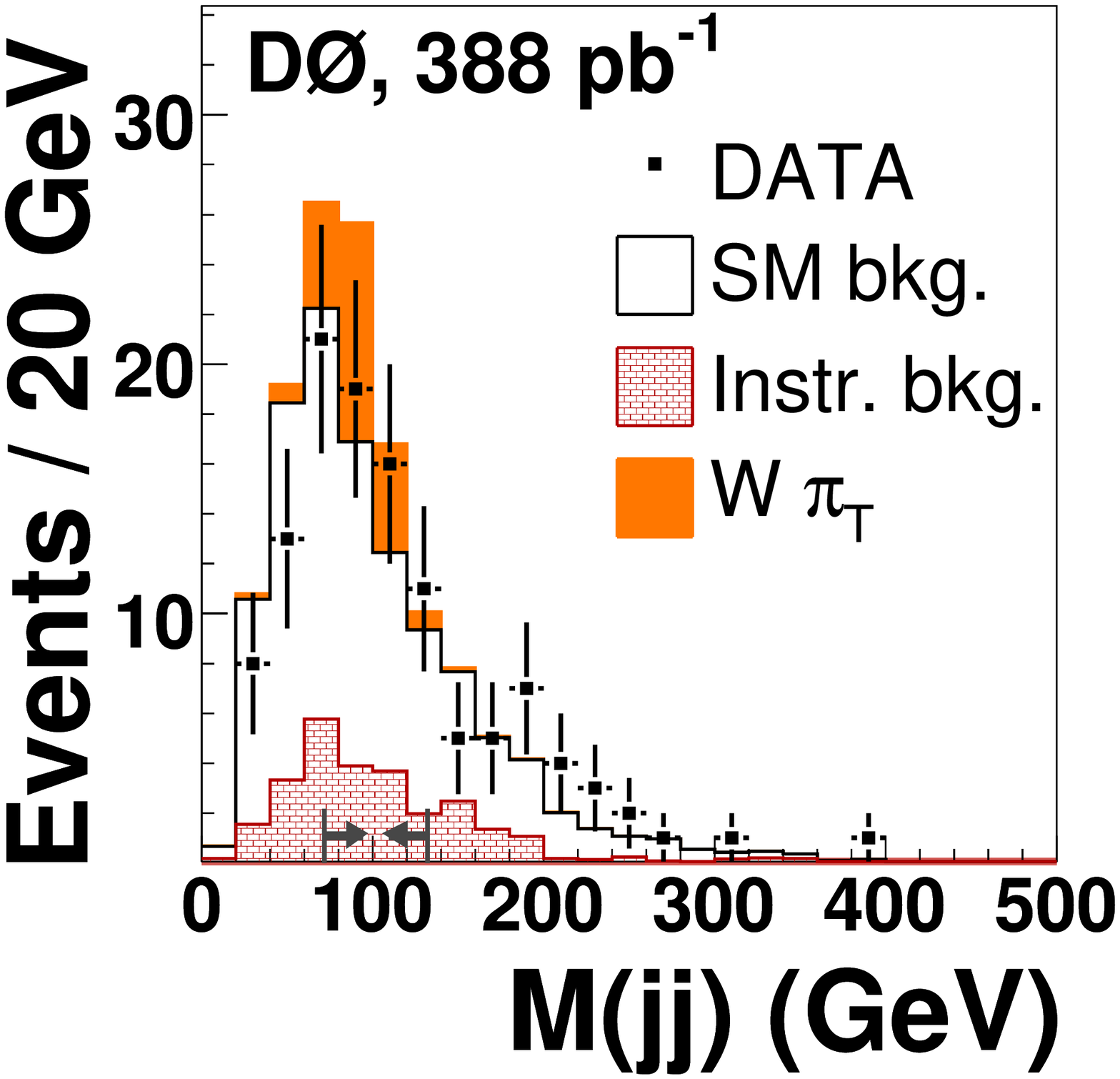}
\includegraphics[scale=0.21]{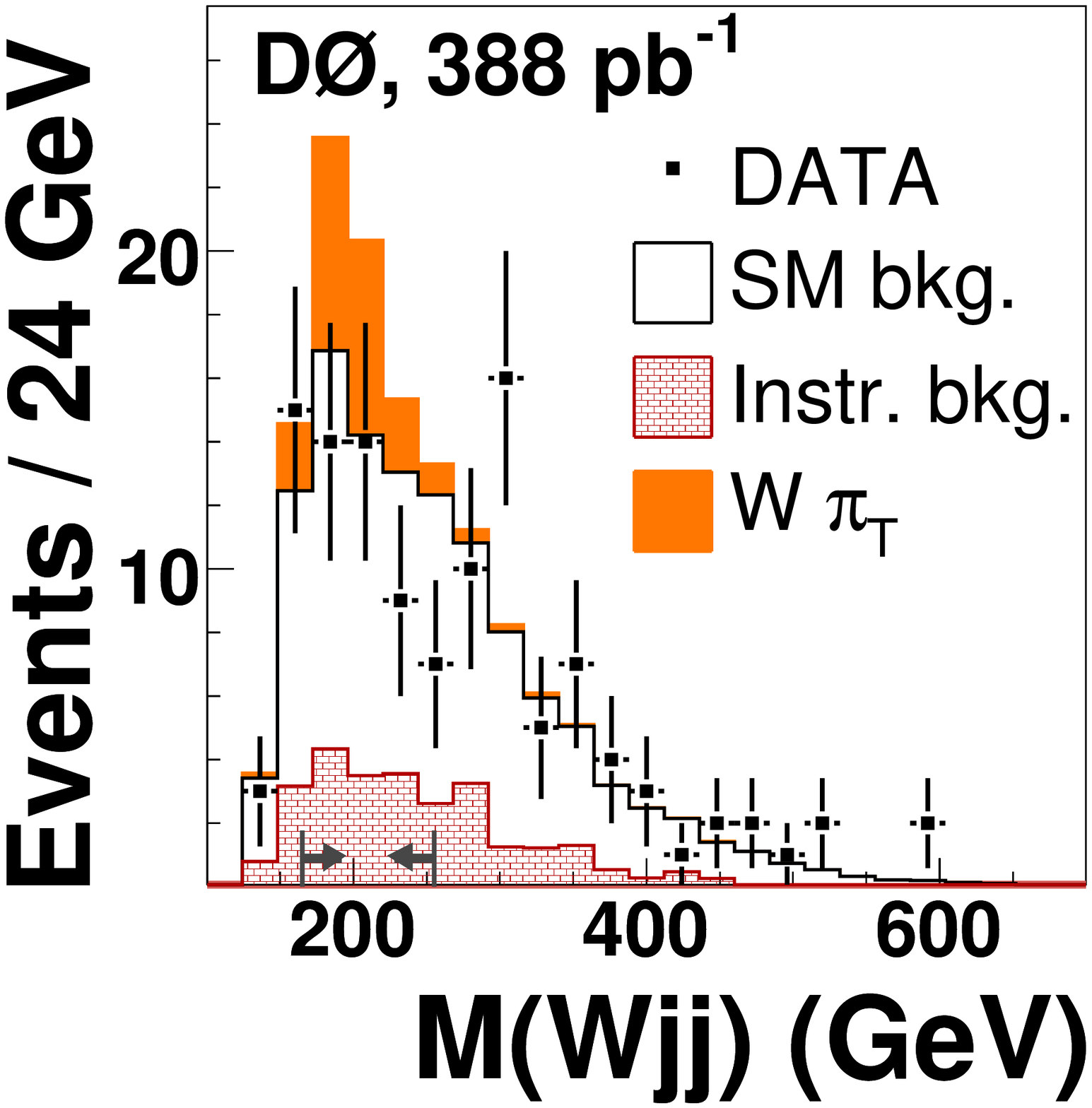}
\caption{\label{fig:wide2} Distributions of $M(jj)$ and $M(Wjj)$ after final kinematic selection. The $W \pi_{T}$ signal is shown for $M(\rho_T)=210$~GeV and $M(\pi_{T})=110$~GeV. Arrows at
the bottom indicate the cuts applied in the cut-based analysis for the signal mass point shown.}
\end{figure}

At this point our data sample is still dominated by background. We
therefore use additional variables that characterize the topology of the
events to discriminate between signal and background. These variables are the
azimuthal angle difference between the two jets $\Delta \phi(j,j)$, the azimuthal
angle difference between the electron and the $\mpt$, $\Delta \phi(e,\mpt)$,
the transverse momentum of the dijet system $p_T(jj)$, the scalar sum of the
transverse momenta of the electron and the two jets $H_T^e$, the invariant
mass of the dijet system $M(jj)$, and the invariant mass of the $W$
boson-dijet system $M(Wjj)$. The technicolor particles are expected to have
narrow widths ($\approx 1$~GeV). We should therefore see enhancements in the
distributions of $M(jj)$ and $M(Wjj)$, consistent in width with the detector
resolution.  $M(jj)$ corresponds to the reconstructed $\pi_T$ mass and
$M(Wjj)$ corresponds to the reconstructed $\rho_T$ mass. We reconstruct the
$W$ boson from the electron and the missing transverse momentum using the $W$ boson 
mass constraint to solve for $p_z$ of the neutrino. If there are two real
solutions, we take the smaller value of neutrino $|p_z|$. If there is only a
complex solution, we take the real part. Distributions of these variables are
shown in Figs.~\ref{fig:wide} and~\ref{fig:wide2}. We use two approaches to
separate signal and background, a cut-based analysis and a neural network
analysis. 

The cut-based analysis is optimized using Monte Carlo simulations to
 maximize the ratio $S/\sqrt{B}$ for every set of technimeson mass values. $S$
 is the expected number of $W\pi_T$ events and $B$ is the expected number of
 background events. For each topological variable, the $S/\sqrt{B}$ ratio is evaluated as a function
of the value of the variable to determine a set of lower, upper, or 
window cuts which maximizes this ratio.

The neural network analysis uses the topological variables
$H_T^e$, $\Delta \phi (e, \mpt)$, $\Delta \phi (jj)$, $p_T(jj)$, the transverse momenta of both jets and 
of the electron and $\mpt$. A two-stage neural network based on the Multi Layer Perceptron algorithm~\cite{NN}
is used. The first stage consists of three independent networks which are trained to reject the three main backgrounds, top quark production, $W+b\bar{b}$ production, and all other $W$+jets production including heavy flavors. Each of these three networks has eight input nodes and one hidden layer with 24 nodes. The second stage network has three input nodes, connected to the outputs of the three networks in the first stage, and one hidden layer with six nodes. 
The second stage network is trained using all nine physics background processes. The networks are trained separately for each set of technicolor mass values. We then apply the trained neural networks to the collider data, technicolor signals, and
 physics and instrumental backgrounds to obtain the discriminator output spectra. We optimize the discriminator cut for every set of techniparticle masses to maximize $S/\sqrt{B}$.

There is no excess in our data over the expected background. We compute upper
limits on the $\rho_T \rightarrow W \pi_T \rightarrow e\nu \; b \bar{b}
(\bar{c})$ production cross section times branching fraction. In the cut-based
analysis, which is a simple counting experiment, we compute an upper 95\% C.L. limit
on the signal using Bayesian statistics~\cite{d0limit}. The neural network analysis
performs a maximum likelihood fit of the data in the $M(\rho_T),M(\pi_T)$
plane to signal and background expectations. The backgrounds are constrained
to their expected values within statistical and systematic
uncertainties. 
The uncertainties in the background event yields total to 10--12\% and the
uncertainty in the signal selection efficiency is 10\% for the cut-based
analysis and 20\% for the neural-net based analysis. 
The largest contributions to the systematic uncertainties are due to  
jet reconstruction efficiency, jet energy scale, $b$-tagging efficiency,
and, only for the signal, from the difference between fast and fully 
simulated detector Monte Carlo. The 95\% C.L. upper
limit on the signal cross section is then determined by the number of signal
events below which lies 95\% of the integral over the resulting likelihood
function.

\begin{figure}[t]
\centering
\includegraphics[scale=0.35]{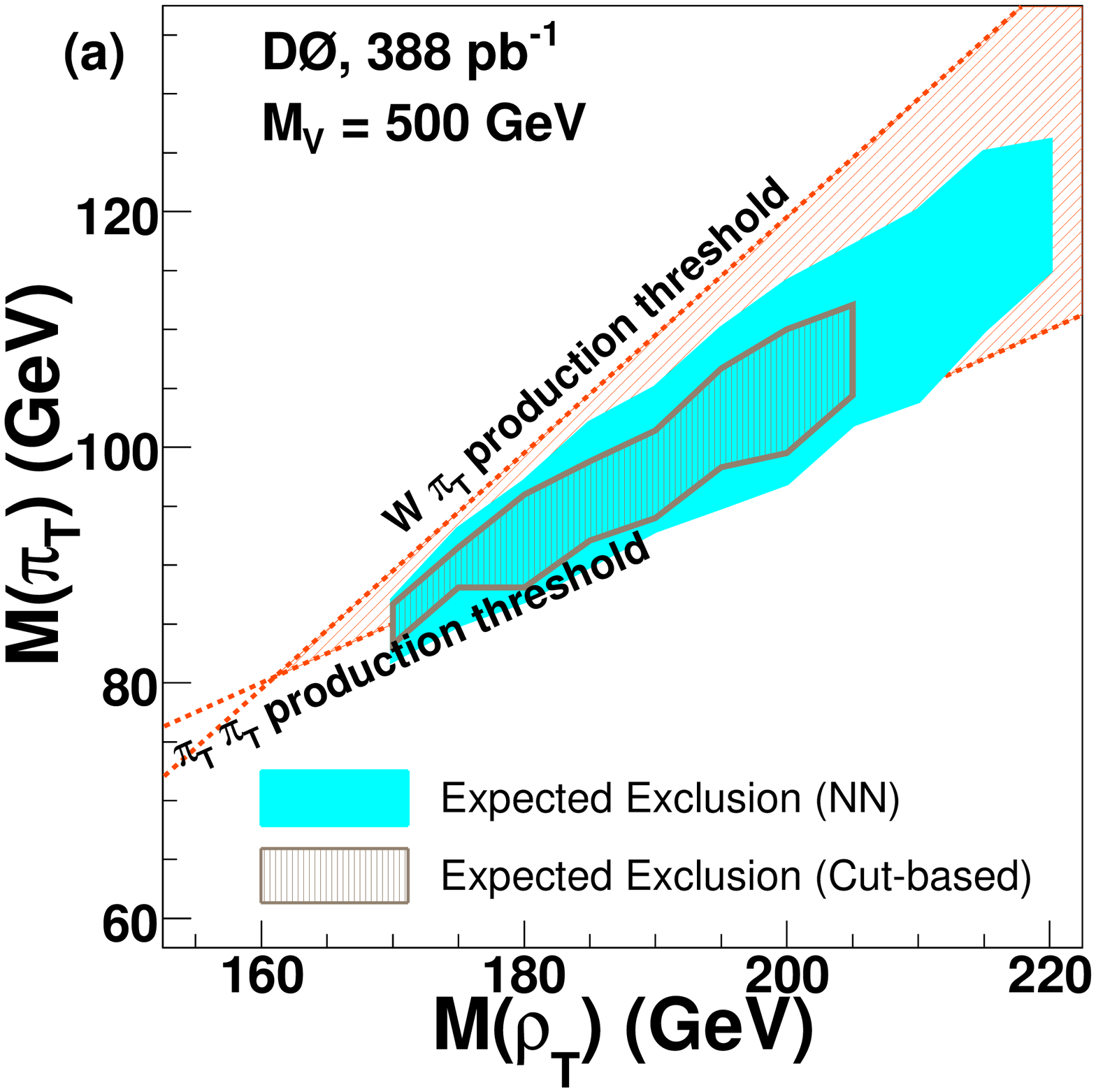}
\includegraphics[scale=0.35]{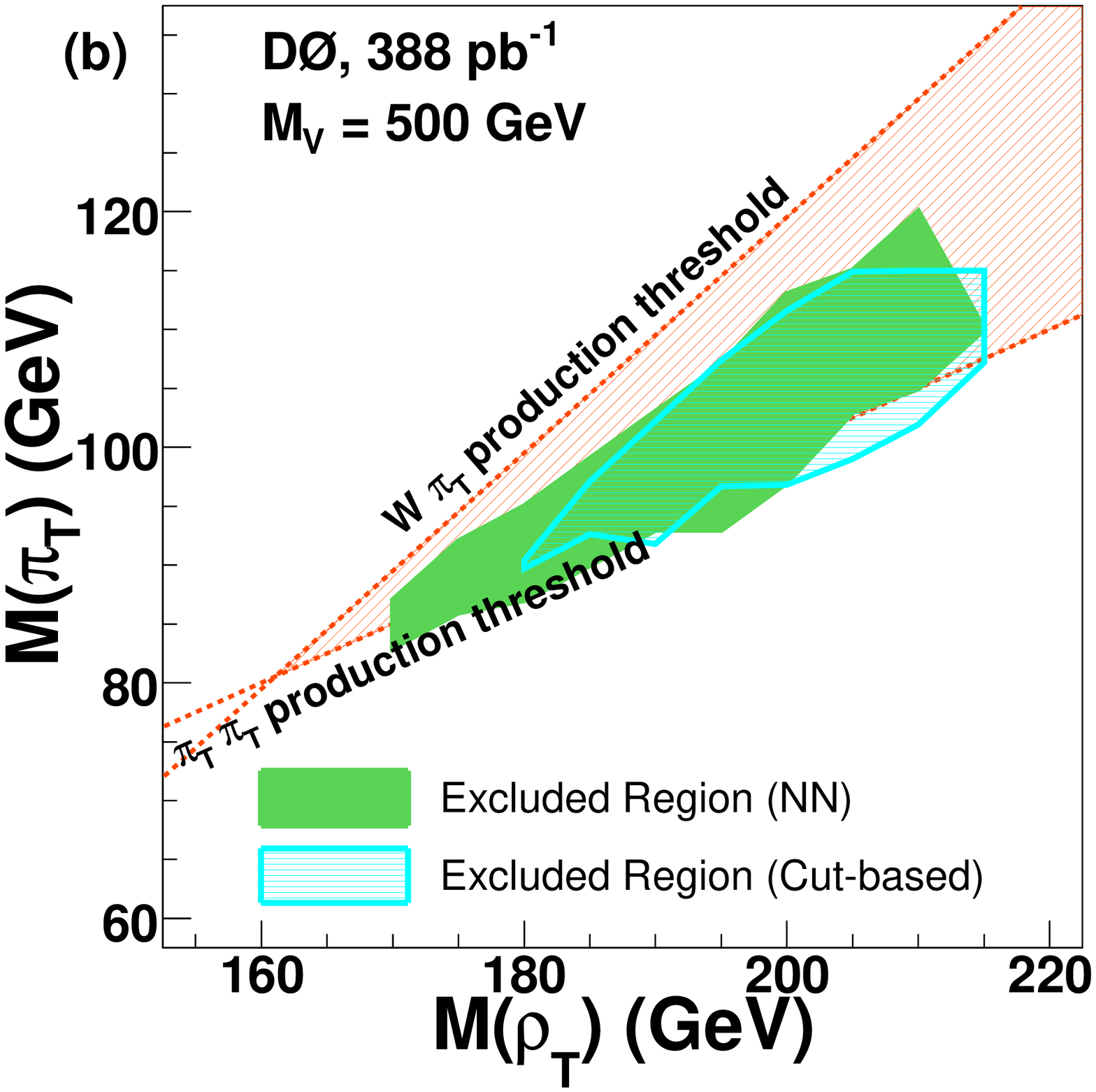}
\caption{Expected  region of exclusion (a) and  excluded region  (b) at the
  95\% C.L. in the $M(\rho_T),M(\pi_T)$ plane  for $\rho_T \rightarrow W
 \pi_T \rightarrow e\nu \; b \bar{b}(\bar{c})$ production with $M_V =
 500$~GeV. Kinematic thresholds from  $W \pi_T$ and  $\pi_T \pi_T$  are shown on the figures.}
\label{fig:limitexp}
\end{figure}

The expected sensitivity and the regions excluded at 95\% C.L. 
by both  analyses in 
the $M(\rho_T),M(\pi_T)$ plane for $M_V=500$~GeV are shown in 
Fig.~\ref{fig:limitexp}.
For $M_V=100$~GeV, only a small region around
$M(\rho_T)=190$~GeV and $M(\pi_T)=95$~GeV can be excluded. 
We note from Fig.~\ref{fig:limitexp}(a), that the expected sensitivity of 
the neural network analysis is
better than that of the cut-based analysis, as indicated by the
larger  95\% C.L. exclusion region.
We quote the  observed 95\% C.L. exclusion region 
in the  $M(\rho_T),M(\pi_T)$ plane in Fig.~\ref{fig:limitexp}(b) by the neural
network analysis as our measurement~\cite{footnote}. 

The results presented in this paper cannot be compared directly to 
those previously published~\cite{PreviousTC}. The CDF
experiment did not use Ref.~\cite{Lane2} and~\cite{Lane4}, 
but rather  the models described in the earlier paper of
Ref.~\cite{Lane1}, a precursor to the TCSM. The LEP
experiments used  Ref.~\cite{Lane2} in which the
cross sections, while appropriate for narrow $\rho_T$ production in $\bar{q}
q$ collisions, are incorrect for off-resonance production in $e^+e^-$
collisions such as at LEP (see Ref.~\cite{Lane3}). 
Although differences in the employed TC models preclude a direct comparison with previous searches, the current search 
achieves a higher sensitivity to the considered physics process.

%
We thank Ken Lane for helpful discussions,
the staffs at Fermilab and collaborating institutions, 
and acknowledge support from the 
DOE and NSF (USA);
CEA and CNRS/IN2P3 (France);
FASI, Rosatom and RFBR (Russia);
CAPES, CNPq, FAPERJ, FAPESP and FUNDUNESP (Brazil);
DAE and DST (India);
Colciencias (Colombia);
CONACyT (Mexico);
KRF and KOSEF (Korea);
CONICET and UBACyT (Argentina);
FOM (The Netherlands);
PPARC (United Kingdom);
MSMT (Czech Republic);
CRC Program, CFI, NSERC and WestGrid Project (Canada);
BMBF and DFG (Germany);
SFI (Ireland);
The Swedish Research Council (Sweden);
Research Corporation;
Alexander von Humboldt Foundation;
and the Marie Curie Program.
%


\begin{thebibliography}{99}
\bibitem{pseudorapidity}
$\eta=-\ln[\tan(\theta/2)]$ and $\theta$ is the polar angle with respect
to the proton beam direction.
\bibitem{footnote}
Consistent scaling of luminosity and background prior to optimization, 
using the new D0 luminosity~\cite{newlumi}
 will lead to somewhat better limits. Nevertheless, 
we choose to keep the analysis consistent with the previous estimate 
of the  luminosity  value~\cite{oldlumi}.
\end{thebibliography}
\end{document}